\documentclass[preprint,showpacs,preprintnumbers,amsmath,amssymb,nofootinbib]{revtex4}
\usepackage{graphicx}

\usepackage{dcolumn}
\usepackage{bm}

\begin{document}
\title{Modulator noise suppression in the LISA Time-Delay
  Interferometric combinations}

\author{Massimo Tinto}
\email{Massimo.Tinto@jpl.nasa.gov}
\affiliation{Jet Propulsion Laboratory, California Institute of
             Technology, Pasadena, CA 91109}

\author{J. W. Armstrong}
\email{John.W.Armstrong@jpl.nasa.gov} 
\affiliation{Jet Propulsion Laboratory, California Institute of
             Technology, Pasadena, CA 91109}

\author{Frank B. Estabrook}
\email{Frank.B.Estabrook@jpl.nasa.gov} 
\affiliation{Jet Propulsion Laboratory, California Institute of
             Technology, Pasadena, CA 91109}

\date{\today}

\begin{abstract}
  LISA (Laser Interferometer Space Antenna) is a mission to detect and
  study low-frequency cosmic gravitational radiation through its
  influence on the phases of six modulated laser beams exchanged
  between three remote spacecraft.  We previously showed how the
  measurements of some eighteen time series of relative frequency or
  phase shifts could be combined (1) to cancel the phase noise of the
  lasers, (2) to cancel the Doppler fluctuations due to non-inertial
  motions of the six optical benches, and (3) to remove the phase
  noise of the onboard reference oscillators required to track the
  photodetector fringes, all the while preserving signals from passing
  gravitational waves. Here we analyze the effect of the additional
  noise due to the optical modulators used for removing the phase
  fluctuations of the onboard reference oscillators. We use the
  recently measured noise spectrum of an individual modulator
  (Klipstein {\it et al.} \cite{Klip06}) to quantify the contribution
  of modulator noise to the first and second-generation Time-Delay
  Interferometric (TDI) combinations as a function of the modulation
  frequency. We show that modulator noise can be made smaller than the
  expected proof-mass acceleration and optical-path noises if the
  modulation frequencies are larger than $\approx 682$ MHz in the case
  of the unequal-arm Michelson TDI combination $X_1$, $\approx 1.08$
  GHz for the Sagnac TDI combination $\alpha_1$, and $\approx 706$ MHz
  for the symmetrical Sagnac TDI combination $\zeta_1$.  These
  modulation frequencies are substantially smaller than previously
  estimated and may lead to less stringent requirements on the LISA's
  oscillator noise calibration subsystem.  The measurements in
  \cite{Klip06} were performed in a laboratory experiment for a range
  of modulation frequencies, but we emphasize that, for the reference
  oscillator noise calibration algorithm to work, the modulation
  frequencies must be equal to the frequencies of the reference
  oscillators.
\end{abstract}

\maketitle

\section{Introduction}
\label{intro}

LISA (Laser Interferometer Space Antenna) is a three-spacecraft deep
space mission, jointly proposed to the National Aeronautics and Space
Administration (NASA) and the European Space Agency (ESA). It will
detect and study low-frequency cosmic gravitational radiation by
observing frequency shifts of laser beams interchanged between
drag-free spacecraft \cite{PPA98}.

Modeling each spacecraft with two optical benches, carrying
independent lasers, beam splitters and photodetectors, we previously
analyzed the measured eighteen time series of frequency shifts (six
one-way laser carrier beams between spacecraft pairs, six between the
two optical benches on each of the three spacecraft, and six more by
over-imposing phase modulations on the laser carrier beams between
spacecraft pairs to monitor rates of the reference oscillators on each
spacecraft).  We showed that there exist several combinations of these
eighteen observable which cancel the otherwise overwhelming phase
noise of the lasers, and the phase fluctuations due to the
non-inertial motions of the six optical benches, and also allow the
removal of the phase noise from the onboard reference oscillators (or
USOs -- "Ultra-Stable Oscillators") required to track the
photodetector fringes, while leaving signals due to passing
gravitational waves \cite{TEA02}.

The analysis in our previous work assumed that noise due to the
electro-optic modulators used to implement the onboard reference
oscillator noise calibration algorithm was negligible.  Here we amend
those results to a more realistic LISA operational configuration where
the effects of the phase fluctuations of the modulators are explicitly
included in the various time-delay interferometric (TDI,
\cite{TD_Living05}) combinations. An experimental investigation for
estimating the magnitude of the phase noise introduced by a
commercially available electro-optical modulator (EOM) on the phase
measurement was recently performed by Klipstein {\it et al.}
\cite{Klip06}. They showed that a modulation frequency of about $8$
GHz would be adequate to suppress EOM noise in a one-way phase
measurement to smaller than a budgeted noise level.  Their conclusion
was conservative in that it did not take account of the transfer
functions of modulator noise to the TDI combinations.  Here we derive
the EOM noise transfer functions and show that smaller modulation
frequencies than those previously identified could be used, implying
less stringent phase noise requirements on the EOMs and perhaps
simpler subsystem design.
 
In Section \ref{LISA_TDI} we give a brief summary of TDI and its
implementation for LISA, and show that the expressions derived in
\cite{TEA02} for removing the Ultra-Stable Oscillator noise from the
TDI combinations, valid for a stationary LISA configuration (so called
``first-generation TDI''), can easily be generalized to the
``second-generation'' TDI combinations (i.e. those accounting for the
the motions of the three spacecraft with respect to each other and
around the Sun).

In Section \ref{Modulator_Spectrum} we derive the transfer functions
of the modulator noises to the second-generation TDI combinations
$X_1$ (unequal-arm Michelson), $\alpha_1$ (Sagnac), and $\zeta_1$
(symmetrized Sagnac).  Using the measurements in \cite{Klip06} on a
particular EOM, we show that selecting modulation frequencies greater
than $\approx 682$ MHz (for $X_1$), $\approx 1.08$ GHz (for
$\alpha_1$), and $\approx 706$ MHz (for $\zeta_1$) results in the
power spectral density of the modulator noise being smaller than the
optical-path and proof-mass noises of these TDI combinations.

\section{Time-Delay Interferometry}
\label{LISA_TDI}

Equal-arm Michelson interferometer detectors of gravitational waves
can observe gravitational radiation by canceling the much larger frequency
fluctuations of the laser light injected into their arms. This is
done by comparing phases of split beams propagated along the equal
(but non-parallel) arms of the detector. The laser frequency
fluctuations affecting the two beams experience the same delay within
the two equal-length arms and cancel out at the photodetector where
relative phases are measured. In this way gravitational wave signals of
dimensionless amplitude less than $10^{-22}$ can be observed
using lasers whose frequency stability can be as large as 
$10^{-13}$.

If the arms of the interferometer have different lengths, as will
inevitably be the case for space-based detectors such as LISA, simple
differencing of the phases on the photodetector does not exactly
cancel the laser phase fluctuations, $p (t)$.  The larger the
difference between the two arms, the larger will be the magnitude of
the laser phase fluctuations affecting the detector response.  If
$L_1$ and $L_2$ are the lengths of the two arms, the laser relative
phase fluctuations remaining in the response is equal to (units in
which the speed of light $c = 1$)
\begin{equation}
\Delta p (t) = p(t - 2L_1) - p(t - 2L_2) \ .
\label{DC}
\end{equation}
In the case of LISA, whose lasers are expected to display relative
frequency fluctuations equal to about $10^{-13}/\sqrt{Hz}$ in the
milliHertz band, and whose arms will differ by a few percent
\cite{PPA98}, equation (\ref{DC}) implies the following expression
for the amplitude of the Fourier components of the uncanceled laser
frequency fluctuations (an over imposed tilde denotes the operation of
Fourier transform)
\begin{equation} 
|{\widetilde {\Delta p}} (\omega)| \simeq |{\widetilde {p}} (\omega)| \ 
2 \ \omega |(L_1 - L_2)| \ .
\label{FDC}
\end{equation}
At $\omega/2\pi = 10^{-3}$ Hz, for instance, and assuming $|L_1 - L_2| \simeq
0.5 \ \ {\rm sec}$, the uncanceled fluctuations from the laser are
equal to $6.3 \times 10^{-16}/\sqrt{\rm Hz}$. Since the LISA
sensitivity goal is about $10^{-20}/\sqrt{\rm Hz}$ in this part of the
frequency band, it is clear that an alternative experimental approach
for canceling the laser frequency fluctuations is needed. 

The solution of this problem is achieved by first ``de-coupling'' the
two arms with the implementation of a multi-photo-receiver design in
which, at each optical bench, the phase difference between the light
entering the arm and the one exiting it is measured, time-tagged, and
digitally recorded.  By then properly time-shifting and linearly
combining these phase measurements exact cancellation of the laser
phase noise is again achieved.  The sensitivity to gravitational radiation
for the practical LISA case is essentially equivalent to that of an
equal-arm Michelson detector. For a physical and historical
description of this technique, called Time-Delay Interferometry (TDI),
the reader is referred to \cite{TD_Living05} and references therein.

\begin{figure}
  \begin{center}
    \includegraphics[width=2.5in, angle = -90.0]{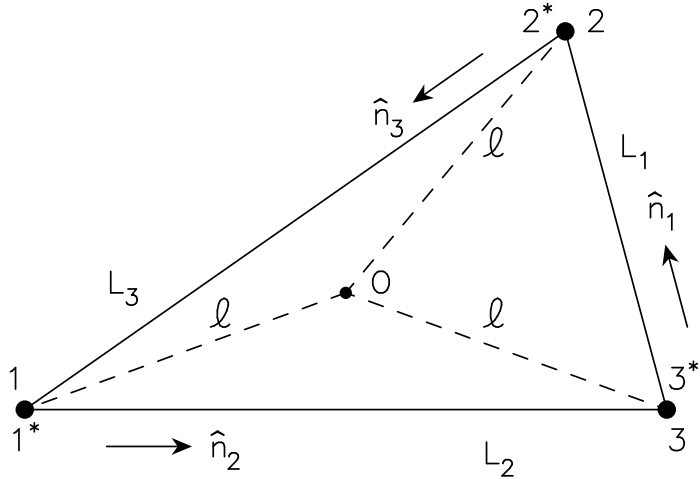}
    \end{center}

    \caption{Schematic LISA configuration.  Each spacecraft is equidistant
  from the point O, in the plane of the spacecraft.  Unit vectors
  $\hat n_i$ point between spacecraft pairs with the indicated
  orientation.  At each vertex spacecraft there are two optical
  benches (denoted 1, $1^*$, etc.), as indicated.}
\label{Fig1}
\end{figure}

In the case of LISA there are six beams exchanged between the three
spacecraft, with the six relative phase measurements $s_{ij}$ ($i,j =
1, 2, 3$) recorded when each received beam is mixed with light from an
independent laser on the receiving optical bench (that laser also
being used to transmit light back along the same arm to the distant
spacecraft{\footnote{This is the simplest configuration; for technical
    reasons, schemes in which two or more of the lasers are remotely
    phase-locked are being considered by the LISA project.  TDI works
    as well for these alternate measurement approaches also
    \cite{TSSA03}.}}).  The frequency fluctuations from all the six
lasers, which enter in each of the six Doppler measurements, must be
removed to levels smaller than those of the secondary (proof mass and
optical path) noises \cite{TEA02} in order to detect and study
gravitational radiation at the predicted amplitudes (see Figure
\ref{Fig1} for a description of the LISA geometry).

As the three spacecraft forming the LISA array orbit the Sun, their
systematic relative motions result in Doppler shifts of the laser
frequencies. These offsets may be as large as about $\pm 20 \ {\rm
  MHz}$ and are measured at the receiving spacecraft via heterodyning.
To limit the offset frequencies to this range either the lasers
must be referenced to an atomic line (such as molecular iodine
stabilization \cite{LC06}) or a sophisticated locking scheme must be
employed \cite{TSSA03}. 

A heterodyne measurement is ``base-banded'' by using a properly
selected tracking frequency generated by a reference oscillator (the
USO) using servo loop feedback. Although this procedure allows us to
accurately track the phases of the photodetector fringes, it
introduces into the phase measurements a residual noise due to the USO
itself \cite{TEA02}.  This USO noise is not negligible and a technique
for removing it from the TDI combinations was devised (see
\cite{TEA02} for more details).  This technique requires the
modulation of the laser beams exchanged by the spacecraft, and the
further measurement of six more inter-spacecraft relative phases by
comparing the sidebands of the received beam against sidebands of the
transmitted beam \cite{TEA02}.

\begin{figure}
\centering
\includegraphics[width=3.0 in, angle=0.0]{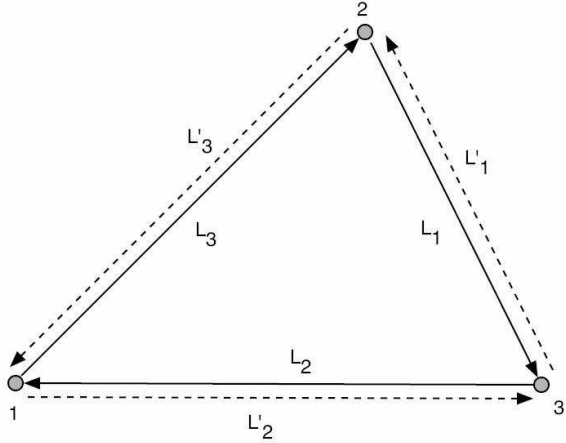}
\caption{Schematic diagram of the LISA configuration involving
  six laser beams. Optical path delays taken in the counter-clockwise
  sense are denoted with a prime, while unprimed delays are in the
  clockwise sense.
\label{Fig2}}
\end{figure}

The time-keeping of the three spacecraft must be synchronized in an
inertial frame, usually taken to be the solar system barycentric
frame.  On each link, the up- and downlink delay times used in the TDI
combinations thus differ due to large relativistic aberrations. This
is first accommodated in the analysis by so called ``modified
generation'' TDI combinations, and in ``second generation'' TDI where
more general ``flexing'' of the LISA configuration is allowed.
Following \cite{TEA04}, the arms are labeled with single numbers given
by the opposite spacecraft; e.g., arm $2$ (or $2^{'}$) is opposite
spacecraft $2$, where primed delays are used to distinguish
light-times taken in the counter-clockwise sense and unprimed delays
for the clockwise light times (see Figure (\ref{Fig2})).  Also the
following labeling convention of the relative phase data will be used.
Explicitly: $s_{23}$ is the one-way phase shift measured at spacecraft
$3$, coming from spacecraft $2$, along arm $1$.  Similarly, $s_{32}$
is the phase shift measured on arrival at spacecraft $2$ along arm
$1'$ of a signal transmitted from spacecraft $3$.  Due to aberration
and the relative motion between spacecraft, $L_1 \neq L_1^{'}$ in
general.  As in \cite{TEA02}, we denote six further data streams
$\tau_{ij}$ ($i,j = 1, 2, 3$), as the intra-spacecraft metrology data
used to monitor the motion of the two optical benches and the relative
phase fluctuations of the two lasers on each of the three spacecraft.

The frequency fluctuations introduced by the lasers, by the optical
benches, by the proof masses, by the fiber optics, by the USOs, and by
the measurements themselves at the photo-detectors (i.e.\ the
shot-noise fluctuations) enter the Doppler observables $s_{ij}$,
$\tau_{ij}$ with specific time signatures; see 
\cite{ETA00,TEA02} for a detailed discussion.  The contribution
$s^\mathrm{GW}_{ij}$ due to GW signals was derived in
\cite{AET99} in the case of a stationary array, and further extended
to the realistic configuration \cite{KTV04} of the LISA array orbiting
the Sun.

The four data streams recorded at spacecraft 1, including Doppler
effects, independent lasers, gravitational wave signals, optical path
noises, proof-mass and bench noises, fiber optics and USO phase fluctuations, are
now given by the following expressions \cite{TEA02,TEA04}
\begin{eqnarray}
s_{31} & = &  \left[\nu_{3} \ (1 - \dot L_2) - \nu^*_{1} - a_{31} \ f_{1} \right] t
+ p_{3,2} - p^{*}_1 - a_{31} \ q_1 - \nu_{3} \ {\hat n_2} \cdot {\vec \Delta_{3,2}} 
\nonumber \\
& & 
+ \nu_{3} \ (1 - \dot L_2) \ \left[ 2 {\hat n_2} \cdot {\vec \delta^{*}_1}  - 
{\hat n_2} \cdot {\vec \Delta^{*}_1} \right] 
+ s^{\rm gw}_{31} + s^{\rm opt. \ path}_{31}
\label{eq:1}
\\
\tau_{31} & = &  \left[\nu_{1} - \nu^*_{1} - c_{31} \ f_{1} \right] t +
p_{1} - p^{*}_1 - c_{31} \ q_1
+ 2 \ \nu_{1} \ {\hat n_3} \cdot ({\vec \delta_1} - {\vec \Delta_1}) + \mu_1 
\label{eq:2}
\\
s_{21} & = & \left[\nu^*_{2} \ (1 - \dot L_{3'}) - \nu_{1} - a_{21} \ f_{1} \right] t
+ p^*_{2,3'} - p_1 - a_{21} \ q_1 + \nu^*_{2} \ {\hat n_3} \cdot {\vec \Delta^*_{2,3'}} 
\nonumber 
\\
& & 
+ \nu^*_{2} \ (1 - \dot L_{3'}) \ \left[ - \ 2 {\hat n_3} \cdot {\vec \delta_1} +
{\hat n_3} \cdot {\vec \Delta_1} \right]
+ s^{\rm gw}_{21} + s^{\rm opt. \ path}_{21}
\label{eq:3}
\\
\tau_{21} & = & \left[\nu^*_{1} - \nu_{1} - c_{21} \ f_{1} \right] t +
p^{*}_{1} - p_1 - c_{21} \ q_1
- \ 2 \ \nu^*_{1} \ {\hat n_2} \cdot ({\vec \delta^*_1} - {\vec \Delta^*_1}) + \mu_1 \ .
\label{eq:4}
\end{eqnarray}
\noindent
For all the down conversions at spacecraft $1$, the USO-generated
frequency $f_1$ is used, and it is multiplied by coefficients
$a_{21}$, $a_{31}$, $c_{21}$, and $c_{31}$ that are the result of
phase-lock loops driving numerically controlled oscillators to track
the large ($ \simeq 20 \ {\rm MHz}$ center frequency) offsets in the
heterodyne phase measurements (the ``beat notes'') \cite{TEA02}.  Thus
the values of these coefficients are determined by the following
expressions
\begin{eqnarray}
a_{31} & = & \frac{\nu_{3} \ (1 - \dot L_2) - \nu^*_{1}}{f_{1}} \ ,
\label{eq:5}
\end{eqnarray}

\begin{eqnarray}
a_{21} & = & \frac{\nu^*_{2} \ (1 - \dot L_{3'}) - \nu_{1}}{f_{1}} \ ,
\label{eq:6}
\end{eqnarray}

\begin{eqnarray}
c_{21}  = - c_{31} & = & \frac{\nu^*_{1} - \nu_{1}}{f_{1}}  \ .
\label{eq:7}
\end{eqnarray}
Eight other relations, for the readouts at vertices $2$ and $3$, are
given by cyclic permutation of the indices in equations
(\ref{eq:1})-(\ref{eq:7}). 

Recent experimental results have indicated that onboard lasers can be
effectively stabilized by referencing their frequency to that of
molecular-line transitions such as those defined by molecular iodine
($I_2$).  Besides providing a high reference frequency stability for
the LISA onboard lasers, $I_2$ stabilization has the advantage of
making the center frequencies of the lasers onboard each spacecraft
essentially equal \cite{LC06}. In the following we will assume the
onboard lasers to be iodine-stabilized, and we will correspondingly
set the coefficients $c_{ij}$ defined in equation (\ref{eq:7}) to be
identically equal to zero.

The gravitational wave phase signal components in equations
(\ref{eq:1}) and (\ref{eq:3}) are given by integrating with respect to
time the equations (1), (2) of reference \cite{AET99}, which related
frequency shifts to metric perturbations. It is these that LISA is
designed to measure at levels set by the optical path phase noise
contributions, $s^{\rm opt. \ path}_{ij}$, due mainly to shot noise
from the low signal-to-noise ratio (SNR) in the links connecting the
distant spacecraft.  The $\tau_{ij}$ measurements will be made with
high SNR so that for them the shot noise is negligible. The other
unavoidable secondary phase noise is due to proof mass
non-gravitational perturbations, described here by the random
displacement vectors ${\vec \Delta_{i}}$ and
${\vec \Delta^{*}_{i}}$.

The expressions of the phase measurements given by equations
(\ref{eq:1} - \ref{eq:4}), when substituted into the second generation
laser-noise-free combinations yield data that, although free of laser
and motional phase noises, are now affected additionally by the USO
phase fluctuations, which have been denoted $q_i$ ($i = 1, 2, 3$) in
equations (\ref{eq:1} - \ref{eq:4}).  For instance, with a state-of-the-art
USO displaying a frequency stability of about $1.0 \times 10^{-13}$ in
the milliHertz frequency band, the corresponding relative frequency
fluctuations, $\dot q_i/\nu_i$, introduced in the laser-noise-free
data combinations would be equal to about $3.0 \times 10^{-20}$,
several orders of magnitude above the secondary noises and LISA
sensitivity goals \cite{PPA98}.

The expressions of the gravitational wave signal, the USO and the
secondary noise sources entering into $X_1$ will in general be
different from those entering into $X$, the corresponding ``first
generation'' unequal-arm Michelson observable derived under the
assumption of a stationary LISA array \cite{AET99,ETA00}.  However,
the magnitude of the corrections introduced by the motion of the array
on the gravitational wave signal, the USO and the secondary noise
sources entering into $X_1$ will all be proportional to the product of
their time derivatives, the spacecraft relative velocities, and the
LISA arm-length.  At $1$ Hz, for instance, the larger corrections (due
to aberration) will be about five orders of magnitude smaller than the
main terms.  Since the amplitude of these corrections scale linearly
with the Fourier frequency, we can completely disregard the time- and
direction-dependence of the time-delays entering into these noise
sources over the entire LISA band \cite{TEA04}.

These considerations imply that the second generation TDI expressions
for the gravitational wave signal, the USOs and the secondary noises
can be expressed in terms of the corresponding first generation TDIs.
For instance, the second generation unequal-arm Michelson combination,
$X_{1q}$, (where the $q$-index indicates the inclusion of the USO
noises) can be written in terms of the corresponding first generation
unequal-arm Michelson combination, $X_{q} (t)$, in the following
manner \cite{TL04}
\begin{equation}
X_{1q} (t) = X_{q} (t) - X_{q} (t - 2L_2  - 2L_3) \ ,
\label{X1fromX}
\end{equation}
in which the time-dependence of the light-travel times can be
completely disregarded. Equation (\ref{X1fromX}) implies that the USO
calibration procedure for the second generation TDI combinations can
actually be evaluated by simply considering the corresponding calibration
expressions derived for the first generation TDI expressions given in
\cite{TEA02}. For this reason, from now on we will focus our attention
on the first generation combinations.

\section{Magnitude of the modulator noise into the TDI combinations}
\label{Modulator_Spectrum}

In the USO calibration scheme first proposed by Bender {\it et al.}
\cite{PPA98}, a second frequency is superimposed on each of the six
main laser beams: specifically, beams originating at spacecraft $i$
are modulated at the frequency $f_i$ generated by its onboard USO. The
main carrier signal, together with a side-band (of intensity perhaps
ten times lower than the intensity in the carrier \cite{PPA98}) are
transmitted, and at the receiving spacecraft $j$ they are heterodyned
at a photo detector with the local laser beam $\nu_j$ also carrying
side-bands \cite{TEA02}. If the frequencies of the sidebands are
carefully selected to be larger than the laser frequency offsets (but
to differ from each other by an amount smaller than the operational
bandwidth of the photo detector) then the lowest two difference
frequencies (or difference phase trains) can be distinguished and
measured at the photo detector.  These two difference phase time
series are given respectively by the difference between incoming and
outgoing carriers, and by the difference of the phases of their
side-bands respectively.  They are then independently further tracked
with coefficients $a_{ij}$ and $b_{ij}$ \cite{TEA02,H01}.  This
process provides six additional data records, $s'_{ij}$.  Consider,
for instance, the phase difference between the second signal
transmitted by bench $3$ and the second at the receiving bench $1^*$
\begin{eqnarray}
s'_{31} & = &  \left[(\nu_{3} + f_{3}) \ (1 - \dot L_2) - \nu^*_{1} -
f_{1} - b_{21} \ f_{1} \right] t + p_{3,2} - p^{*}_1 
\nonumber
\\
& + & q_{3,2} - (1 \ + \ b_{21}) \ q_1 + m_{3,2} - m^*_1 
\label{eq:8}
\end{eqnarray}
\noindent
where, for simplicity, we have omitted the terms associated with the
optical bench noises, the optical path and proof-mass noises, and the
contribution from a possibly present gravitational wave signal. Note
that now the expressions for the measurements $s'_{ij}$ include also
the phase noises $m_i, m^*_i \ \ i = 1, 2, 3$ introduced by the
modulators on the measured phase differences $s'_{ij}$. These terms
were neglected in \cite{TEA02} and here we amend that analysis.

Note that the numerical coefficient $b_{21}$, determined by the
following equation
\begin{eqnarray}
b_{21} & = & \frac{(\nu_{3} + f_{3}) \ (1 - \dot L_2) - (\nu^*_{1} + f_{1})}{f_{1}} \ ,
\label{eq:9}
\end{eqnarray}
is distinct from $a_{21}$ given by equation (\ref{eq:5}) (although
they will be close if all the $f_i$ are close).

The quantities $r^{(m)}_{21} \equiv (s_{21} - s'_{21})/f_3$ and
similarly $r^{(m)}_{31} \equiv (s_{31} - s'_{31})/f_2$ (and cyclic
permutations of their indices) enter into the algorithm presented in
\cite{TEA02} for removing the USO noises from the various TDI
combinations. Now they explicitly show their dependence on the
modulator noises $m_i, m^*_i \ \ i = 1, 2, 3$, and are related to the
$r_{ij}$ expressions introduced in \cite{TEA02} (which did not include
the modulator noises) by the following expressions
\begin{eqnarray}
r^{(m)}_{31} & = & r_{31} - \frac{m_{3,2} - m^*_1}{f_3} 
\nonumber
\\
r^{(m)}_{21} & = & r_{21} - \frac{m^*_{2,3} - m_1}{f_2} 
\label{r_func}
\end{eqnarray}
with the others obtained as usual via permutation of the spacecraft
indices. 

\subsection{Modulator Noise in the $X$-combination}

As a result of the presence of modulator noises in the $s'_{ij}$
measurements (and consequently in the combinations $r^{(m)}_{ij}$) the
first-generation TDI unequal-arm Michelson combination given in
equation (27) of \cite{TEA02} (which now will include combinations of
the $r^{(m)}_{ij}$ measurements) will be affected by the modulator
noises. If we denote with $X^{(m)}$ the resulting new unequal-arm
Michelson combination, it is easy to see that this can be written as
the sum of $X$ (as given in \cite{TEA02}) with terms due to the
modulator noises. The resulting expression is equal to
\begin{eqnarray}
X^{(m)} & = & X - 
a_{12} \ f_2  \
\left[
\frac{m_{1,22} - m^*_{2,322}}{f_2} 
+
\frac{m^*_{2,3} - m_{1}}{f_2} 
+
\frac{m^*_{1} - m_{3,2}}{f_3} 
+
\frac{m_{3,2} - m^*_{1,22}}{f_1} 
\right]
\nonumber
\\
& + & 
a_{13} \ f_3 \
\left[
\frac{m^*_{1,33} - m_{3,233}}{f_3} 
+
\frac{m_{3,2} - m^*_{1}}{f_3} 
+
\frac{m_{1} - m^*_{2,3}}{f_2} 
+
\frac{m^*_{2,3} - m_{1,33}}{f_1} 
\right]
\nonumber
\\
& - & 
a_{21} \ f_1 \
\left[
\frac{m^*_{1} - m_{3,2}}{f_3} 
+
\frac{m_{3,2} - m^*_{1,22}}{f_1} 
\right]
+
a_{31} \ f_1 \ 
\left[
\frac{m_{1} - m^*_{2,3}}{f_2} 
+
\frac{m^*_{2,3} - m_{1,33}}{f_1} 
\right] \ .
\label{Xm}
\end{eqnarray}
The expression above leads to the following estimation of the spectral
density of the noise, $S_{X^{(m)}}$, in the TDI combination $X^{(m)}$
\begin{eqnarray}
S_{X^{(m)}} & = & S_{X} \ + \ 4 \sin^2(\omega L) \ 
\left[
S_{m_1} (a_{12} + a_{13} + a_{31})^2 + 
S_{m^*_1} (a_{12} + a_{13} + a_{21})^2
\right.
\nonumber
\\
& & \left.
+ S_{m^*_2} \ a^2_{12} + S_{m_3} \ a^2_{13} 
\right] \ ,
\label{Sxm}
\end{eqnarray}
which has been written as the sum of the secondary noise spectra
$S_{X}$ of the combination $X$ (not containing the modulator noises)
and the spectra of the modulator noises themselves. Note that for a
figure of merit equation (\ref{Sxm}) conveniently assumes all USOs to
have equal frequency $f$, and all the armlengths to be equal to a
common value $L$.

Assuming a maximum beat-note frequency offset of $20$ MHz in each arm,
the coefficients $a_{ij}$ can be taken all equal to $a \equiv 20 \ 
{\rm MHz} / f$, where $f$ is the USO or modulation
frequency{\footnote{In the experimental setup used by \cite{Klip06} to
    characterize an individual EOM's noise it was not necessary that
    the modulation frequency be the same as the USO frequency.  We
    emphasize that in LISA's actual USO calibration subsystem the
    modulation frequency must equal the USO frequency; see Figure 5 of
    \cite{TEA02}.}.  Given the measured modulator noise spectrum
  (figure 6 in \cite{Klip06}), from equation (\ref{Sxm}) it is then
  possible to determine the value of the USO frequency which would
  make the modulator noise contribution smaller than the secondary
  noises (\cite{ETA00}, Section IV) affecting the $X$ combination over
  the entire LISA band. This can be done my maximizing over the
  angular frequency $\omega$ the following function (see equation
  \ref{Sxm} above)
\begin{equation}
\sin(\omega L) \ \sqrt{80 \ \frac{S_m (\omega)}{S_X
    (\omega)}} \ \times 20 \ {\rm MHz} \ ,
\label{Xfmod}
\end{equation}
where $S_m (\omega)$ is the spectrum of the noise associated with one
modulator.  Over the frequency band relevant to LISA, the measured
$S_m (\omega)$ for a single EOM can be approximated by the following analytic
expression (see figure 6 in \cite{Klip06})

\begin{eqnarray}
S_m (\omega) & = & \ 
2.8 \ \times 10^{-9} \ \omega^{-1} \ \ \ {\rm cycles}^2 \ {\rm Hz}^{-1} 
\ \ \ {\rm when:} 
\ \ \ 10^{-4} \le \omega/2\pi \le 8 \times 10^{-3} \ {\rm Hz} \nonumber \\ 
& = & \ 5.5 \ \times 10^{-8} \ \ \ \ \ \ \ \ \ {\rm cycles}^2 \ {\rm Hz}^{-1} 
\ \ \ {\rm when:} \ \ \ 
8 \times 10^{-3} \le \omega/2\pi \le 3 \times 10^{-2} \ {\rm Hz} \nonumber \\
& = & \ 3.8 \ \times 10^{-10} \ \omega^{-3} \ \ {\rm cycles}^2 \ {\rm Hz}^{-1} 
\ \ \ {\rm  when:} \ \ \ 
3 \times 10^{-2} \le \omega/2\pi \le 5 \times 10^{-1} \ {\rm Hz} \nonumber \\
& = & \ 1.2 \ \times 10^{-11} \ \ \ \ \ \ \ \ {\rm cycles}^2 \ {\rm Hz}^{-1} 
\ \ \ {\rm when:} \ \ \ 
5 \times 10^{-1} \le \omega/2\pi \le 1 \ \ {\rm Hz} \ .
\label{Sm}
\end{eqnarray}
Note that the expressions of the power spectral densities of the
noises entering into the $X$, $\alpha$, and $\zeta$ combinations are
given in Section IV of \cite{ETA00} as fractional frequency
fluctuations, and therefore need to be converted to the same units as
$S_m$ before being used for direct comparison.

After maximizing the function given in equation (\ref{Xfmod}) we
conclude that the modulation frequency $f$ should be equal to or
larger than $682$ MHz for the modulator noise level in the
$X$-combination to be smaller than the remaining secondary noises.
A numerical comparison of modulation noise in $X$ at this level,
modulation noise if $f = 8 \ {\rm GHz}$ is adopted, and the combined
secondary noises is shown in Fig. \ref{Fig3}.

\begin{figure}
\centering
\includegraphics[width=5.0in]{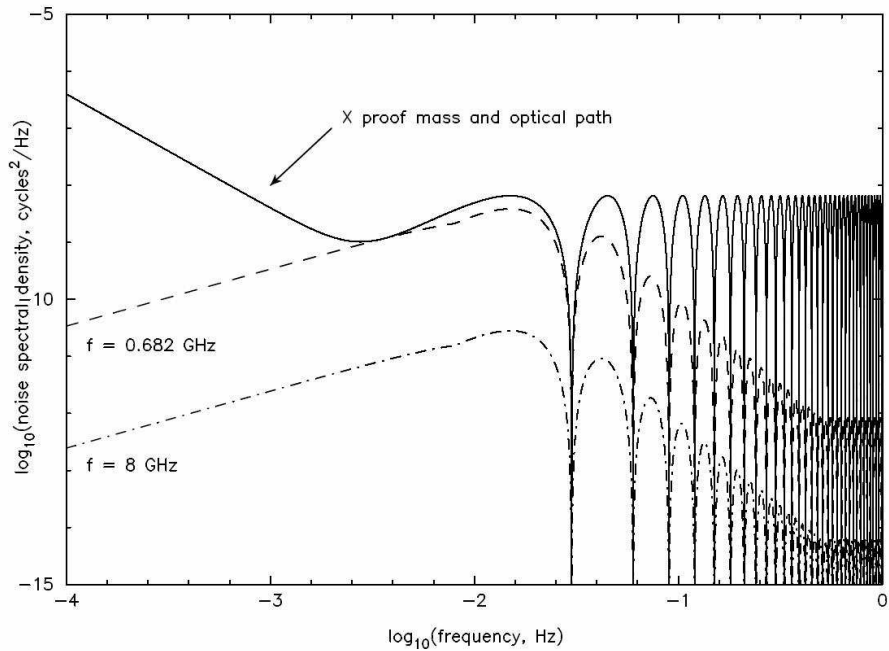}
\caption{Numerical comparison of modulation noise in $X$ with a
  modulation frequency $f$ equal to $682 \ {\rm MHz}$
  (dashed-line), $ 8 \ {\rm GHz}$ (dash-dot line), and the combined
  secondary noises (solid line).}
\label{Fig3}
\end{figure}

\subsection{Modulator Noise in the $\alpha$-combination}

If we denote with $\alpha^{(m)}$ the resulting new Sagnac
combination, it is easy to see that this can be written as the
sum of $\alpha$ (as given in \cite{TEA02}) and terms
due to the modulator noises. The resulting expression is equal to 
\begin{eqnarray}
\alpha^{(m)} & = & \alpha + 
[a_{32} \ f_2  + a_{13} \ f_3]
\
\left[
\frac{m_{1} - m^*_{2,3}}{f_2} 
\right]
-
[a_{23} \ f_3  + a_{12} \ f_2]
\left[
\frac{m^*_{1} - m_{3,2}}{f_3} 
\right]
\nonumber
\\
& - &
a_{12} \ f_2
\left[
\frac{m^*_{3,2} - m_{2,12}}{f_2} 
\right]
+
a_{13} \ f_3
\left[
\frac{m_{2,3} - m^*_{3,13}}{f_3} 
\right]
\label{alpham}
\end{eqnarray}
The expression above leads to the following estimation of the spectral
density of the noise, $S_{\alpha^{(m)}}$, in the TDI combination $\alpha^{(m)}$
\begin{eqnarray}
S_{\alpha^{(m)}} & = & S_{\alpha} 
\ + \ 
(a_{32} + a_{13})^2 \ 
(S_{m_1} + S_{m^*_2}) 
\ + \ 
(a_{23} + a_{12})^2 \ 
(S_{m^*_1} + S_{m_3})
\nonumber
\\
& + &
|(a_{13} + a_{12} \ e^{ i \omega L})|^2 
\ S_{m_2}
\ + \
|(a_{12} + a_{13} \ e^{i \omega L})|^2 
\ S_{m^*_3}
\label{Salpham}
\end{eqnarray}
which has been written as the sum of the noise spectra of the
modulator noise-free combination $\alpha$ and the contribution to the
overall spectrum coming from the modulator noise. All the armlengths
have been taken again to be equal to a common value $L$.

Again taking the maximum beat-note frequency offset to be $20$ MHz in
each arm, the coefficients $a_{ij}$ can be treated as all equal to $a
\equiv 20 \ {\rm MHz} / f$. Now the function that needs to be
maximized to determine the value of the USO frequency which
makes the modulator noise smaller than the LISA secondary noises
over the entire LISA band is equal to
\begin{equation}
\sqrt{\frac{S_m (\omega)}{S_\alpha (\omega)} \ (20 +
 4  \cos(\omega L))} \times 20 \ {\rm MHz} \ .
\label{alphafmod}
\end{equation}
By calculating the maximum value of the above function over the LISA
operational band we conclude that a modulation frequency
$f$ equal to or larger than $\approx 1.08$ GHz will suppress the
modulator noise in the $\alpha$-combination to a level smaller
than the secondary noises. A numerical comparison of the resulting
modulation noise in $\alpha$ with the secondary noises is shown in
Fig. \ref{Fig4}.

\begin{figure}
\centering
\includegraphics[width=5.0in]{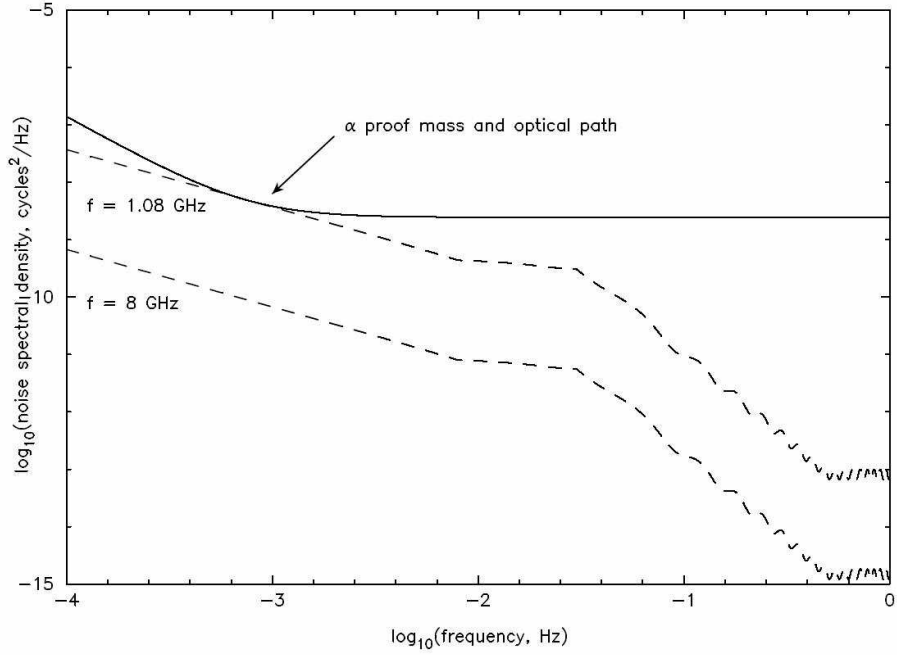}
\caption{Numerical comparison of modulation noise in $\alpha$ with a
  modulation frequency $f$ equal to $1.08 \ {\rm GHz}$ (dashed-line),
  $ 8 \ {\rm GHz}$ (dash-dot line), and the combined secondary noises
  (solid line).}
\label{Fig4}
\end{figure}

\subsection{Modulator Noise in the $\zeta$-combination}

We denote with $\zeta^{(m)}$ the new symmetrized Sagnac combination,
and write it again as the sum of $\zeta$ (as given in \cite{TEA02})
with added terms due to the modulator noises. The resulting expression is
equal to
\begin{eqnarray}
\zeta^{(m)} & = & \zeta + 
\frac{f_1}{3} (a_{21} - a_{31}) 
\
\left[
\left(\frac{m^*_{3,3} - m_{2,13}}{f_2} \right)
-
\left(\frac{m_{1,1} - m^*_{2,31}}{f_2} \right)
+
\left(\frac{m_{2,2} - m^*_{3,12}}{f_3} \right)
-
\left(\frac{m^*_{1,1} - m_{3,21}}{f_3} \right)
\right]
\nonumber
\\
& + & 
\frac{f_2}{3} (a_{32} - a_{12})
\
\left[
\left(\frac{m^*_{1,1} - m_{3,21}}{f_3} \right)
-
\left(\frac{m_{2,2} - m^*_{3,12}}{f_3} \right)
+
\left(\frac{m_{3,3} - m^*_{1,23}}{f_1} \right)
-
\left(\frac{m^*_{2,2} - m_{1,32}}{f_1} \right)
\right]
\nonumber
\\
& + &
\frac{f_3}{3} (a_{13} - a_{23}) 
\
\left[
\left(\frac{m^*_{2,2} - m_{1,32}}{f_1} \right)
-
\left(\frac{m_{3,3} - m^*_{1,23}}{f_1} \right)
+
\left(\frac{m_{1,1} - m^*_{2,31}}{f_2} \right)
-
\left(\frac{m^*_{3,3} - m_{2,13}}{f_2} \right)
\right] \ .
\label{zetam}
\end{eqnarray}
In order to identify the minimum modulation frequency $f$ that
suppresses the modulator noise below the secondary noises, it is
useful to set $a_{ij} = a_{ji}$, which is allowed if there is no
flexing and
\begin{figure}
\centering
\includegraphics[width=5.0in]{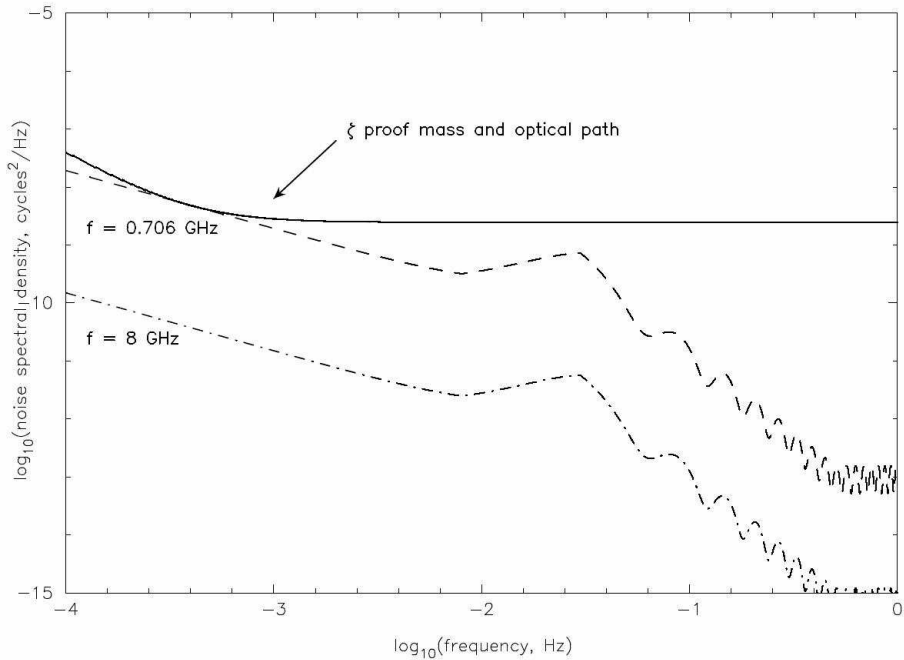}
\caption{Numerical comparison of modulation noise in $\zeta$ with a
  modulation frequency $f$ equal to $706 \ {\rm MHz}$ (dashed-line),
  $ 8 \ {\rm GHz}$ (dash-dot line), and the combined secondary noises
  (solid line).}
\label{Fig5}
\end{figure}
the six lasers are stabilized to the same $I_2$ transition frequency.
Again assuming a maximum beat-note frequency offset of $20$ MHz in
each arm, the coefficients $a_{ij}$ can be taken as $a \equiv 20 \ 
{\rm MHz} / f$. A careful analysis shows that the function which needs
to be evaluated to identify the value of the USO frequency $f$
required to make the modulator noise smaller than the LISA secondary
noises is 

\begin{equation}
\sqrt{\frac{16 \ [2 - \cos(2 \pi f L)] \ S_m (\omega)}{3 \ S_\zeta (\omega)}} \times 20 \ {\rm MHz} \ .
\label{zetafmod}
\end{equation}
By calculating the maximum value of the function above over the LISA
band we conclude that a modulation frequency $f$ equal to or larger
than $706$ MHz will suppress the modulator noise in $\zeta$ to a level
smaller than the secondary noises. A numerical comparison of the
resulting modulation noise in $\zeta$ is shown in Fig. \ref{Fig5}.

\section*{Acknowledgments}

We thank Bill Klipstein, Daniel Shaddock, and Brent Ware for several
stimulating discussions about modulator noise.  This research was
performed at the Jet Propulsion Laboratory, California Institute of
Technology, under contract with the National Aeronautics and Space
Administration.  Massimo Tinto and John Armstrong were supported under
research task 05-BEFS05-0014. Frank B. Estabrook is a Distinguished
Visiting Scientist at the Jet Propulsion Laboratory.


\begin{references}
  
\bibitem{Klip06} W.Klipstein, P.G. Halverson, R. Spero, R. Cruz, \& D.
  Shaddock. In: {\it Proceedings of the 6th International LISA
    symposium}, Editor(s): S.M. Merkowitz and J.C. Livas, AIP
  Conference Proceedings Volume 873, Greenbelt, Maryland (USA), 19-23
  June 2006. ISBN: 978-0-7354-0372-7.

\bibitem{PPA98} P. L. Bender, K. Danzmann, and the LISA Study Team,
{\it Laser Interferometer Space Antenna for the Detection of
    Gravitational Waves, Pre-Phase A Report}, Doc. MPQ 233
  (Max-Planck-Instit\"ut f\"ur Quantenoptik, Garching, 1998).

\bibitem{TEA02} M. Tinto, F.B. Estabrook, \& J.W. Armstrong {\it
    Phys. Rev. D}, {\bf 65}, 082003 (2002).

\bibitem{TD_Living05} M. Tinto, and S.V.Dhurandhar, 
{\it Living Reviews in Relativity}, {\bf 8}, 4 (2005).

\bibitem{TSSA03} M. Tinto, D.A. Shaddock, J. Sylvestre, \& J.W. Armstrong {\it
    Phys. Rev. D}, {\bf 67}, 122003 (2003).

\bibitem{LC06} V. Leonhardt \& J.B. Camp {\it Appl. Opt.}, {\bf 45},
  4142, (2006)

\bibitem{TEA04} M. Tinto, F.B. Estabrook,  \& J.W. Armstrong {\it
    Phys. Rev. D}, {\bf 69}, 082001 (2004).

\bibitem{ETA00} F.B. Estabrook, M. Tinto, \& J.W. Armstrong {\it
    Phys. Rev. D}, {\bf 62}, 042002 (2000).

\bibitem{AET99}   J.W. Armstrong, F.B. Estabrook, \& M. Tinto {\it
    Ap. J}, {\bf 527}, 814 (1999).

\bibitem{KTV04} A. Krolak, M. Tinto, \& M. Vallisneri {\it
    Phys. Rev. D}, {\bf 70}, 022003 (2004).

\bibitem{STEA03} D.A. Shaddock, M. Tinto, F.B. Estabrook, \& J.W. Armstrong {\it
    Phys. Rev. D}, {\bf 68}, 061303 (2003).
  
\bibitem{CH03} N.J. Cornish \& R.W. Hellings {\it Class. Quantum
    Grav.}, {\bf 20}, 4851 (2003).

\bibitem{TL04} M. Tinto, \& S.L. Larson {\it
    Phys. Rev. D}, {\bf 70}, 062002 (2004).

\bibitem{7} R.W. Hellings, G. Giampieri, L. Maleki, M. Tinto,
  K. Danzmann, J. Homes, \& D. Robertson, {\it Optics Communications},
  {\bf 124}, 313, (1996).

\bibitem{H01} R.W. Hellings, {\it Phys. Rev. D}, {\bf 64}, 022002 (2001).
  
\end{references}
\end{document}